# Microscopic dynamics of collective acoustic excitations in simple liquids

Yixin Xu[1], Xing Xiang[1], Zhigang Li[1*] and Yanguang Zhou[1*]

*[1]Department of Mechanical and Aerospace Engineering, The Hong Kong University of Science and Technology, Clear Water Bay, Kowloon, Hong Kong SAR*

## Abstract

In this letter, we systematically investigate the microscopic dynamics of collective vibrational excitations in simple liquids. The thermodynamic states of simple liquids are unified to the mean atomic free volume. Our results show that longitudinal acoustic collective vibrational excitations are always observed in the simple liquids even when the liquids are viscous, in which the atomic free volume is larger than the cross point of the corresponding mean propagation length and the atomic diffusion limit. This is because some long-wavelength longitudinal acoustic collective vibrational excitations can still propagate in viscous liquids. However, transverse acoustic collective vibrational excitations in viscous liquids become localized since both short- and long-wavelength transverse acoustic collective vibrational excitations have propagation lengths smaller than the atomic diffusion limit. Therefore, transverse acoustic collective vibrational excitations may not be detected in simple liquids. The propagation length of macroscopic elastic and shear waves which are the mechanical response of long-wavelength longitudinal and transverse collective vibrational excitations, respectively, is further calculated to quickly determine the propagation-to-localization crossover of collective vibrational excitations in simple liquids. Our findings here advance the understanding of microscopic dynamics of collective vibrational excitations in simple liquids.

---

*Author to whom all correspondence should be addressed. Email: mezli@ust.hk (Z. Li); maeygzhou@ust.hk (Y. Zhou)

Collective vibrational excitations (CVEs) are fundamental for understanding many of the physical properties of liquids, such as dynamical excitations, mechanical and thermal transport properties [1–5]. For instance, the relaxation of transverse CVEs in liquids is found to determine the corresponding thermodynamic crossover from elastic to viscous [6]. In long-wavelength limits, CVEs in liquids may be described using the classical elasticity theory, in which the system is treated as a continuum with macroscopic elastic constants. However, the local microscopic structure of the system and the interatomic force dominantly determine the nature of these CVEs when their wavelength is comparable to the mean interatomic distance.

In 1946, Frenkel [7] suggested that CVEs in liquids can be understood by the existing "solid-like" features, e.g., the atomic motions can be projected into the oscillating diagram at their instantaneous equilibrium positions. Based on this conception, CVEs in simple liquids, including both longitude and transverse modes, have been consistently observed through inelastic scattering experiments [8–11] and atomistic simulations [4,12–14]. While CVEs are observed in liquids, their microscopic origin and their relations to macroscopic elastic and shear waves remain largely unexplored, leading to a gap between the macroscopic response and the microscopic dynamics in liquids. In this letter, we present atomistic investigations to delineate the microscopic origins of CVEs in simple liquids. We find that the existence of CVEs in simple liquids is determined by the mean atomic free volume which unifies all thermodynamic states of simple liquids.

All molecular dynamic simulations to calculate the viscoelastic and CVEs in simple liquids are implemented using the Large-scale Atomic/Molecular Massively Parallel Simulator (LAMMPS) [15]. The simple liquids are described by a 12-6 Lennard-Jones (LJ) potential via $U(r) = 4\varepsilon \left[ (\sigma/r)^{12} - (\sigma/r)^{6} \right]$ [16], where $\varepsilon$ and $\sigma$ are the potential energy and particle diameter parameters, respectively, and $r$ is the interatomic distance. The detailed parameters used in the LJ potential are given in the Supplemental Material [17]. All the results in the following are expressed in LJ-reduced units, where $\varepsilon$, $\sigma$, and $m$ are measures of energy, distance, and mass, respectively.

The viscosity $\eta$ of simple LJ liquids is then calculated using the Green-Kubo method [18,19] through $\eta = \dfrac{V}{T} \lim\limits_{t \to \infty} \displaystyle\int_0^t \langle p_{ij}(\tau) p_{ij}(0) \rangle d\tau$, in which $\tau$, $V$, $T$, $p_{ij}$ is autocorrelation time, the system volume, temperature, and the nondiagonal elements of pressure tensor, respectively. The bulk elasticity $\vartheta$ is calculated via $\vartheta = \dfrac{V}{T} \lim\limits_{t \to \infty} \displaystyle\int_0^t \langle \delta p_{ii}(\tau) \delta p_{ii}(0) \rangle d\tau$ where $\delta p_{ii}(\tau) = p_{ii}(\tau) - \langle p_{ii} \rangle$ and $p_{ii}(\tau)$ is the diagonal terms of the pressure tensor of the system. We also introduce the instantaneous viscosity $\eta^\infty$ and bulk elasticity $\vartheta^\infty$ of the system to identify the relaxation of systems (see discussions below). The atomic mean free volume $V_f$ [20] is computed based on the density $\rho$ of the system, i.e., $V_f = 1/\rho - V_c$ where $V_c$ denotes the intrinsic cage volume of atoms when the dynamics of systems are frozen [21,22]. Furthermore, we demonstrate that $V_f$ generalizes the diffusion coefficient $D$ calculated via $D = \dfrac{1}{6} \dfrac{d \langle |\vec{u}(t)|^2 \rangle}{dt}$, where $\langle |\vec{u}(t)|^2 \rangle$ is the mean square displacement, as well as the viscosity $\eta$ [17]. More calculation details about the free volume model can be found in Ref. [17]. Meanwhile, we calculate the current density function to quantify CVEs in the simple liquids through the current density correlation function $C(\vec{q}, t) = \dfrac{1}{N} \langle \vec{J}(\vec{q}, t) \cdot \vec{J}(-\vec{q}, 0) \rangle$ [23], in which $\vec{q}$, $t$ and $\vec{J}$ are wave vectors, time, and current density function, respectively. The current density function $\vec{J}$ is calculated via $\vec{J}(\vec{q},t) = \sum_{i}^{N} \vec{v}_i(t) e^{\mathrm{i}\vec{q}\cdot\vec{r}_i(t)}$, where $\vec{v}_i$ is the atomic velocity. As expected, CVEs in LJ liquids are clearly observed in our calculated current density function spectrums (**Figures 1a** and **1b**). We further fit the current density function with a damped harmonic oscillator (DHO) function through

$$C(|\vec{q}|,\omega) = A\frac{2\Gamma}{(\omega^2 - \omega_0^2)^2 + (\Gamma\omega)^2},$$ where $\omega_0$, $\Gamma$ and $A$ are the frequency, damping, and amplitude respectively. The lifetimes of these CVEs are then calculated via $\tau = \frac{2}{\Gamma}$ (**Figure 1c**). The dispersions of these CVEs can be obtained by fitting the possible wave vectors in the systems (**Figure 1d**), which can be used to calculate the corresponding mean group velocity, i.e.,

$$v_g = \frac{\partial\omega}{\partial|\vec{q}|}.$$

Based on the viscoelastic theory of liquid dynamics [3,5,14,24–26], the elasticity holds the longitude CVEs, and the viscosity is responsible for transverse CVEs. In simple liquids, longitude acoustic (LA) CVEs can always be observed because density fluctuations can occur due to a strong repulsive force between atoms when they approach each other [27]. However, whether there are transverse acoustic (TA) CVEs in liquids depends on the shear force in the long spatial range which can be quantified using the corresponding viscosity. Our results show that there always exists LA CVEs in our systems while TA CVEs appear only when the system viscosity $\eta$ is larger than a critical value, i.e., 0.364 [17].

We then investigate the microscopic origin of CVEs in simple liquids. It is known that the viscosity of liquids results from the dynamics of atomic mass transport. The dynamics of atomic mass transport in simple liquids can be viewed as rearrangements of atoms caused by collisions and random diffusions [7], which can be further exemplified by the $V_f$. It is noted that other thermodynamic properties such as the temperature and pressure of the system can also be reduced to this intrinsic volumetric parameter $V_f$. Our results show that the diffusivity and viscosity of the systems can be well described by $V_f$ regardless of the thermodynamic state of the systems [17].

The principal properties of CVEs include group velocity, and relaxation time (or equivalently, the propagation length). Here, we first investigate the relation between $V_f$ and the mean group velocities of CVEs in the systems The mean group velocities of both LA and TA CVEs in the system can be well depicted by $V_f$ (**Figure 2**) through an exponential decay law $v_g = Ae^{-V_f/V_c}$, in which $A$ is a constant and $V_c$ is the critical mean atomic free volume which is of the order of the size of the cage formed around an atom by its closest neighbours. The extracted $V_c$ based on our calculations for TA and LA CVEs are 0.172 and 0.459 (**Figure 2**), respectively. It is found that the mean group velocity of LA CVEs decreases a little bit when $V_f$ is smaller than $V_c$. When $V_f$ becomes larger than $V_c$, the mean group velocity of TA CVEs decreases quickly with $V_f$. This is because the shear force between local pair atoms decays fast when $V_f$ is above $V_c$. However, the repulsive force between atoms is strong even when $V_f$ is over $V_c$ [6]. Therefore, the mean group velocity of LA CVEs decreases much slower than that of TA CVEs, as shown in **Figure 1**.

We next move to the average propagation length $\bar{\Lambda} = v_g \bar{\tau}$ of these CVEs. It is known that the CVEs are assumed to travel as waves when their propagation length is larger than the characteristic length, which may be identified as the minimum interatomic spacing $d$ [28,29]. Our calculated average propagation length of both LA and TA CVEs is much longer than $d$ at a small $V_f$, and decays following $\Lambda = \Lambda_0 e^{-V_f/V_c}$ (**Figure 3a**). For LA and TA CVEs, the average propagation length becomes smaller than $d$ at $V_f = 0.349$ and $V_f = 0.122$, respectively.

However, the minimum interatomic spacing focuses on disordered systems with negligible inter-diffusions, and therefore is usually used as a characteristic length in amorphous solids [30]. In fluids, local bond relaxation induced by atomic diffusion truncates collective vibrations and may lead to a sharp decrease in their relaxation time [7]. Therefore, the characteristic length for liquids should account for atomic diffusion and may not be simplified by the minimum interatomic spacing. It is known that the diffusion limit $u_{limit}$ [17] of atoms represents the upper limit of atomic diffusion distance in the system. Therefore, the diffusion limit $u_{limit}$ of atoms may be used as the characteristic length for liquids. The particle movements are mainly contributed by atomic oscillations (atomic diffusion motion) when the average propagation length is larger (smaller) than $u_{limit}$. These CVEs with a $u_{limit}$ larger than the average propagation length cannot be regarded as collective waves. Furthermore, we may determine the solid-like elastic regime and the hydrodynamic regime in liquids by comparing the $u_{limit}$ and $d$, i.e., 1) $u_{limit} < d$ denotes the atomic displacement is limited by the local bond environment and the corresponding atoms can diffuse to their nearest atomic positions utmost; 2) $u_{limit} > d$ indicates that the atoms lose their bond connectivity and the liquids become hydrodynamic. The atomic free volume to distinguish the solid-like elastic regime and the hydrodynamic regime of liquids can be also calculated based on the Frenkel model [3,7], which is 0.346 for simple liquids studied here and denoted as $V_{Frenkel}$. Our results show that the calculated $V_{Frenkel}$ is close to the value of 0.360 determined by comparing the $u_{limit}$ and $d$.

It should be noted that our calculated average propagation length assumes all the TA or LA CVEs have the same group velocity and relaxation time. In real situations, the CVEs should be wavelength dependent. We further calculate the propagation lengths of LA and TA CVEs at two specific $|\vec{q}|$ points (**Figures 3b** and **3c**), i.e., short-wavelength CVEs at $|\vec{q}| \approx 3.4$ and long-wavelength CVEs at $|\vec{q}| \approx 0.5$. For liquids in the elastic regime whose $V_f$ is smaller than $V_{Frenkel}$, the short-wavelength CVEs can propagate a distance larger than $u_{limit}$ while are truncated within interatomic distance $d$ (**Figure 3b**). This is because the short-wavelength CVEs are easily scattered by the breaking of local static bonds due to the rapid diffusion of particles in the fluids. However, these long-wavelength CVEs generally propagate a distance longer than $d$ and $u_{limit}$ for liquids in the elastic regime (**Figure 3c**). This is because long-wavelength CVEs are primarily influenced by spatial density fluctuations in the fluid, whereas local bond relaxations cause negligible changes in spatial density distributions due to interdiffusion. Meanwhile, the Frenkel model [7] demonstrates that TA CVEs become localized in hydrodynamic regime whose $V_f$ is larger than $V_{Frenkel}$ [31]. For viscous liquids, LA CVEs with short wavelengths start to localize and LA CVEs with long wavelengths remain propagating [31]. Our results show that these LA CVEs in viscous liquids with short wavelengths become localized since $\bar{\Lambda}_{LA} < u_{limit}$ (**Figure 3b**), and these LA vibrations with long wavelengths can still propagate as $\bar{\Lambda}_{LA} > u_{limit}$ (**Figure 3c**). However, the average propagation length of TA CVEs in viscous liquids is always smaller than $u_{limit}$ (**Figures 3b** and **3c**), and therefore the corresponding TA CVEs are localized. Furthermore, the average group velocity of TA CVEs decreases one magnitude order when the $V_f$ of the corresponding liquid reduces to $V_{Frenkel}$ (**Figure 2**), and the corresponding dispersion becomes flat and vague [17]. For LA CVEs in viscous liquids, their average group velocity decreases a little compared to that in elastic liquids (**Figure 2**), and the LA CVEs remain propagating in the systems [17].

It is known these long wavelength LA and TA CVEs determine the elastic and the shear mechanical response [14,24,32], respectively. Here, we also calculate the propagation length of macroscopic elastic and shear waves in the liquids to directly quantify the corresponding elasticity and viscosity. The propagation length of the elastic (shear) waves is calculated as $\varLambda = v\tau$, where $v$ is the wave velocity and is simply assumed as the mean group velocity of CVEs in our system, i.e., $v = v_g$, and $\tau$ is the relaxation time of the elastic (shear) waves. The relaxation time of the elastic and shear waves of the system is calculated via $\tau_\eta = \eta/\eta^\infty$ and $\tau_\vartheta = \vartheta/\vartheta^\infty$, respectively, in which $\eta^\infty$ ($\vartheta^\infty$) is the instantaneous viscous (elastic) response and is calculated by

$$\eta^\infty = \lim_{\tau \to 0} \frac{V}{T} \langle p_{ij}(\tau) p_{ij}(0) \rangle \ (\vartheta^\infty = \lim_{\tau \to 0} \frac{V}{T} \langle \delta p_{ii}(\tau) \delta p_{ii}(0) \rangle)$$ [33]. Our results show that both

the elastic and shear waves decrease with $V_f$ (**Figure 4**), in according to the decay of LA and TA CVEs with $V_f$ (**Figure 3a**), respectively. More specifically, both elastic and shear waves can propagate in the system over a long distance of $10^2 \sim 10^3$ when $V_f < 0.05$ (**Figure 4**). This is because of the long relaxation time of elastic and shear waves (i.e., $\sim 10^2$) at a small system $V_f$ [17]. When the system $V_f$ is smaller than the critical free volume for atomic diffusions (i.e., 0.360 [17]), the atomic diffusion within $V_f$ can be ignored and atomic motions are mainly vibrations (i.e., oscillations). Therefore, these long-wavelength acoustic CVEs can propagate a long distance in the system as shown in **Figure 3c**. When the system $V_f$ is above 0.05, the propagation length of both elastic and shear waves decreases dramatically with $V_f$ (**Figure 4**). This is because the atomic diffusion becomes important with the increase of $V_f$ and the CVEs will be strongly scattered by the atomic diffusion [34–36].

Before closing, we also investigate the propagation properties of the elastic and shear waves. It is shown that the elastic waves can propagate with a distance larger than the atomic diffusion limit (**Figure 4**) when the system $V_f$ is smaller than the cross point (i.e., 0.360) which is identical to $V_{Frenkel}$. As we discussed above, the elastic (shear) wave is mainly the response of LA (TA) long-wavelength CVEs. Long wavelength LA CVEs exist in the system with a $V_f$ smaller than $V_{Frenkel}$ (**Figure 3c**) and therefore the elastic waves can propagate in the system. Long-wavelength TA CVEs may have propagation lengths smaller than the atomic diffusion limit $u_{limit}$ (**Figure 3c**) even when the corresponding $V_f$ of the system is smaller than $V_{Frenkel}$. As a result, the shear waves in the system become localized when the corresponding $V_f$ is above 0.201 (i.e., the cross point) which is lower than $V_{Frenkel}$ (**Figure 4**). Our results also show that whether elastic waves propagate or localize in simple liquids can be well distinguished by the Frenkel model (**Figure 4**), while the propagation property of shear waves in the systems will be overestimated by the Frenkel model (**Figure 4**). This is because the Frenkel line here is extracted from the crossover between atomic oscillation and diffusion, encompassing both longitudinal and transverse modes [17].

In summary, we systematically investigate the microscopic dynamics of CVEs in simple liquids. The thermodynamic states of all simple liquids are unified to the mean atomic free volume $V_f$. Our results show that even if the mean propagation length of LA CVEs becomes smaller than the atomic diffusion limit $u_{limit}$ when the system $V_f$ is over $V_{Frenkel}$, there are still long-wavelength LA CVEs propagate in the systems. The LA CVEs can therefore be always observed in simple liquids. However, TA CVEs become localized when the system $V_f$ is over $V_{Frenkel}$ since both short- and long-wavelength TA CVEs have propagation lengths smaller than $u_{limit}$. The TA

CVEs therefore may not be observed in the liquid with a $V_f$ above $V_{Frenkel}$. Meanwhile, we calculate the propagation length of macroscopic elastic and shear waves which are the mechanical response of long-wavelength LA and TA CVEs, respectively. Our results show that the elastic (shear) waves remain propagating and can be observed when the system $V_f$ is lower than the cross point (i.e., 0.346 for elastic waves and 0.201 for shear waves). As a result, we may also use these macroscopic mechanical waves to roughly examine the propagation-to-localization crossover of CVEs in simple liquids.

**Acknowledgments**

Y. Zhou acknowledges the fund from the Research Grants Council of the Hong Kong Special Administrative Region under Grant C7002-22Y, C6020-22G, 260206023 and 16207124, the fund from the Guangdong Natural Science Foundation under Grant No. 2024A1515011407.

# Reference

[1] R. Nossal, *Collective Motion in Simple Classical Fluids*, Phys. Rev. **166**, 81 (1968).

[2] N. P. Kryuchkov, L. A. Mistryukova, A. V. Sapelkin, V. V. Brazhkin, and S. O. Yurchenko, *Universal Effect of Excitation Dispersion on the Heat Capacity and Gapped States in Fluids*, Phys. Rev. Lett. **125**, 125501 (2020).

[3] K. Trachenko and V. V. Brazhkin, *Collective Modes and Thermodynamics of the Liquid State*, Rep. Prog. Phys. **79**, 016502 (2015).

[4] V. V. Brazhkin and K. Trachenko, *Collective Excitations and Thermodynamics of Disordered State: New Insights into an Old Problem*, J. Phys. Chem. B **118**, 11417 (2014).

[5] D. Bolmatov, *The Phonon Theory of Liquids and Biological Fluids: Developments and Applications*, J. Phys. Chem. Lett. **13**, 7121 (2022).

[6] T. Iwashita, D. M. Nicholson, and T. Egami, *Elementary Excitations and Crossover Phenomenon in Liquids*, Phys. Rev. Lett. **110**, 205504 (2013).

[7] J. Frenkel, *Kinetic Theory of Liquids* (Oxford University Press, 1946).

[8] S. Hosokawa, M. Inui, Y. Kajihara, K. Matsuda, T. Ichitsubo, W.-C. Pilgrim, H. Sinn, L. E. González, D. J. González, S. Tsutsui, et al., *Transverse Acoustic Excitations in Liquid Ga*, Phys. Rev. Lett. **102**, 105502 (2009).

[9] X. Y. Li, H. P. Zhang, S. Lan, D. L. Abernathy, T. Otomo, F. W. Wang, Y. Ren, M. Z. Li, and X.-L. Wang, *Observation of High-Frequency Transverse Phonons in Metallic Glasses*, Phys. Rev. Lett. **124**, 225902 (2020).

[10] E. Pontecorvo, M. Krisch, A. Cunsolo, G. Monaco, A. Mermet, R. Verbeni, F. Sette, and G. Ruocco, *High-Frequency Longitudinal and Transverse Dynamics in Water*, Phys. Rev. E **71**, 011501 (2005).

[11] E. Guarini, F. Barocchi, A. De Francesco, F. Formisano, A. Laloni, U. Bafile, M. Celli, D. Colognesi, R. Magli, A. Cunsolo, et al., *Collective Dynamics of Liquid Deuterium: Neutron Scattering and Approximate Quantum Simulation Methods*, Phys. Rev. B **104**, 174204 (2021).

[12] L. E. Bove, F. Formisano, F. Sacchetti, C. Petrillo, A. Ivanov, B. Dorner, and F. Barocchi, *Vibrational Dynamics of Liquid Gallium at 320 and $970\phantom{\rule{0.3em}{0ex}}\mathrm{K}$*, Phys. Rev. B **71**, 014207 (2005).

[13] X. Wang, A. Bhattacharjee, and S. Hu, *Longitudinal and Transverse Waves in Yukawa Crystals*, Phys. Rev. Lett. **86**, 2569 (2001).

[14] R. M. Khusnutdinoff, C. Cockrell, O. A. Dicks, A. C. S. Jensen, M. D. Le, L. Wang, M. T. Dove, A. V. Mokshin, V. V. Brazhkin, and K. Trachenko, *Collective Modes and Gapped Momentum States in Liquid Ga: Experiment, Theory, and Simulation*, Phys. Rev. B **101**, 214312 (2020).

[15] S. Plimpton, *Fast Parallel Algorithms for Short-Range Molecular Dynamics*, J. Comput. Phys. **117**, 1 (1995).

[16] A. Barbot, M. Lerbinger, A. Hernandez-Garcia, R. García-García, M. L. Falk, D. Vandembroucq, and S. Patinet, *Local Yield Stress Statistics in Model Amorphous Solids*, Phys. Rev. E **97**, 033001 (2018).

[17] *See Supplemental Material for Details of MD Simulations; Free Volume Model; Determine the Frenkel Line; Dynamic Structure Factor; Diffusion Limit and Relaxation Time of Macroscopic Waves.*

[18] M. P. Allen and D. J. Tildesley, *Computer Simulation of Liquids: Second Edition*, 2nd ed. (Oxford University Press, Oxford, 2017).

[19] B. Hess, *Determining the Shear Viscosity of Model Liquids from Molecular Dynamics Simulations*, The Journal of Chemical Physics **116**, 209 (2002).

[20] D. L. Hogenboom, W. Webb, and J. A. Dixon, *Viscosity of Several Liquid Hydrocarbons as a Function of Temperature, Pressure, and Free Volume*, The Journal of Chemical Physics **46**, 2586 (1967).

[21] D. Turnbull and M. H. Cohen, *Free-Volume Model of the Amorphous Phase: Glass Transition*, The Journal of Chemical Physics **34**, 120 (1961).

[22] D. Turnbull and M. H. Cohen, *On the Free-Volume Model of the Liquid-Glass Transition*, The Journal of Chemical Physics **52**, 3038 (1970).

[23] J. P. Boon and S. Yip, *Molecular Hydrodynamics* (Courier Corporation, 1991).

[24] J. Ashwin and A. Sen, *Microscopic Origin of Shear Relaxation in a Model Viscoelastic Liquid*, Phys. Rev. Lett. **114**, 055002 (2015).

[25] K. Trachenko, *Viscosity and Diffusion in Life Processes and Tuning of Fundamental Constants*, Rep. Prog. Phys. **86**, 112601 (2023).

[26] J. Goree, Z. Donkó, and P. Hartmann, *Cutoff Wave Number for Shear Waves and Maxwell Relaxation Time in Yukawa Liquids*, Phys. Rev. E **85**, 066401 (2012).

[27] U. Bafile, E. Guarini, and F. Barocchi, *Collective Acoustic Modes as Renormalized Damped Oscillators: Unified Description of Neutron and x-Ray Scattering Data from Classical Fluids*, Phys. Rev. E **73**, 061203 (2006).

[28] A. Ioffe and A. Regel, *Progress in Semiconductors*, London **4**, 237 (1960).

[29] S. Mukhopadhyay, D. S. Parker, B. C. Sales, A. A. Puretzky, M. A. McGuire, and L. Lindsay, *Two-Channel Model for Ultralow Thermal Conductivity of Crystalline Tl3VSe4*, Science **360**, 1455 (2018).

[30] Y. Luo, X. Yang, T. Feng, J. Wang, and X. Ruan, *Vibrational Hierarchy Leads to Dual-Phonon Transport in Low Thermal Conductivity Crystals*, Nat Commun **11**, 1 (2020).

[31] V. V. Brazhkin, Yu. D. Fomin, A. G. Lyapin, V. N. Ryzhov, and K. Trachenko, *Two Liquid States of Matter: A Dynamic Line on a Phase Diagram*, Phys. Rev. E **85**, 031203 (2012).

[32] V. A. Levashov, *Contribution to Viscosity from the Structural Relaxation via the Atomic Scale Green-Kubo Stress Correlation Function*, The Journal of Chemical Physics **147**, 184502 (2017).

[33] F. Jaeger, O. K. Matar, and E. A. Müller, *Bulk Viscosity of Molecular Fluids*, The Journal of Chemical Physics **148**, 174504 (2018).

[34] Y. Zhou, S. Xiong, X. Zhang, S. Volz, and M. Hu, *Thermal Transport Crossover from Crystalline to Partial-Crystalline Partial-Liquid State*, Nat Commun **9**, 1 (2018).

[35] Y. Zhou and S. Volz, *Thermal Transfer in Amorphous Superionic Li2S*, Phys. Rev. B **103**, 224204 (2021).

[36] B. Wu, Y. Zhou, and M. Hu, *Two-Channel Thermal Transport in Ordered–Disordered Superionic Ag2Te and Its Traditionally Contradictory Enhancement by Nanotwin Boundary*, J. Phys. Chem. Lett. **9**, 5704 (2018).

[37] V. V. Brazhkin, Yu. D. Fomin, A. G. Lyapin, V. N. Ryzhov, E. N. Tsiok, and K. Trachenko, ``Liquid-Gas'' Transition in the Supercritical Region: Fundamental Changes in the Particle Dynamics, Phys. Rev. Lett. **111**, 145901 (2013).

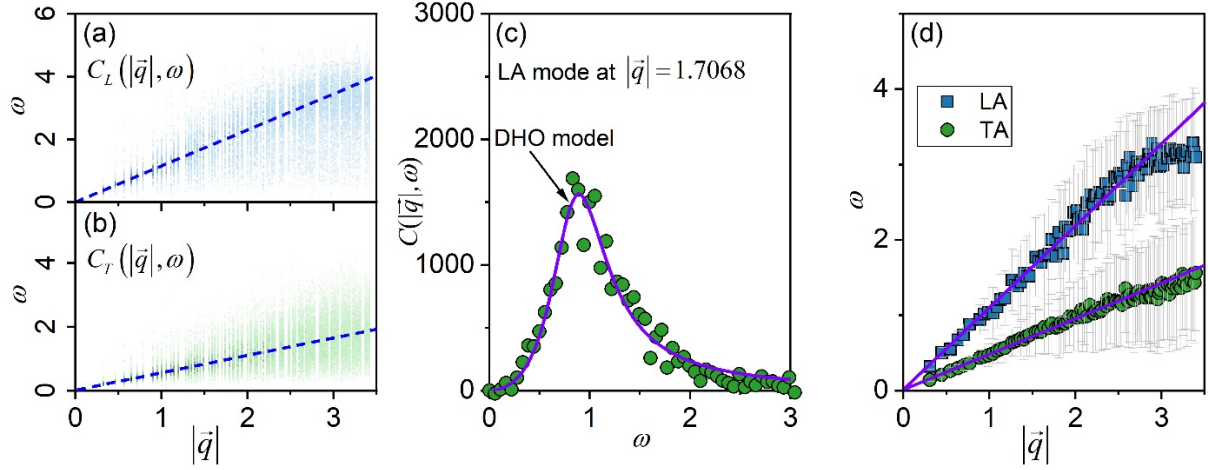

**Figure 1** Current correlation functions of (a) LA and (b) TA collective vibrational excitations in LJ liquids. (c) Fitting the current density with the damped harmonic oscillator (DHO) model at a specific wave vector. (d) The corresponding LA and TA dispersion relations and the error bar denotes the linewidth of these collective vibrational excitations.

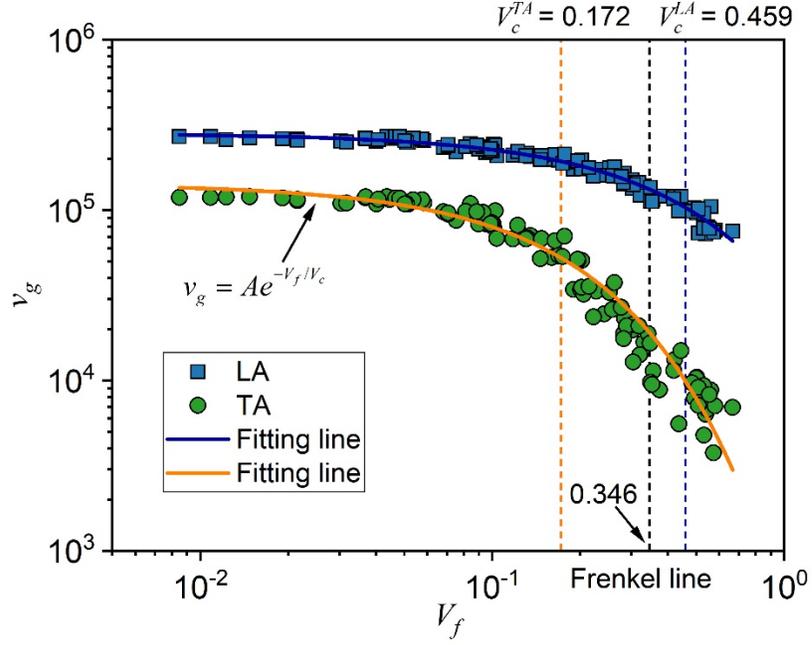

**Figure 2** Group velocities of acoustic collective vibrational excitations versus the atomic mean free volume $V_f$, following an exponential relation of $v_g = Ae^{-V_f/V_c}$ where the decay constant $V_c$ can be understood as the critical volume when collective vibrational excitations are diminished based on the decay $v_g(V_c) = A/e$. The Frenkel line [7] at $V_{Frenkel}$ is calculated based on the evolutions of oscillations of atomic motions [37], and the calculation details can be found in Ref. [17]. The temperature and pressure ranges from $0.172$ to $0.689$ and from $2.45 \times 10^{-4}$ to $7.36 \times 10^{-3}$, respectively.

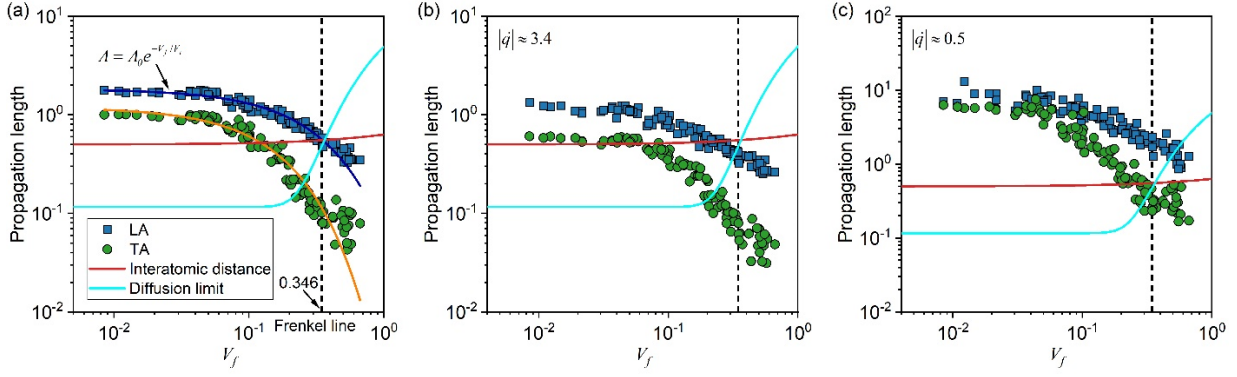

**Figure 3** The propagation length of acoustic collective vibrational excitations versus the atomic mean free volume $V_f$. (a) The average propagation length is calculated as $\bar{\Lambda} = v_g \bar{\tau}$ where $v_g$ is the group velocity and $\bar{\tau}$ is the average relaxation time, i.e., $\bar{\tau} = \frac{1}{N} \sum \tau(\vec{q}, \omega)$. The propagation length decays with $V_f$ following an exponential relation of $\Lambda = \Lambda_0 e^{-V_f/V_c}$. The propagation length of LA and TA collective vibrational excitations with (b) short-wavelength $|\vec{q}| \approx 3.4$ and (c) long-wavelength $|\vec{q}| \approx 0.5$, in which $\Lambda(\vec{q}, \omega) = v_g \tau(\vec{q}, \omega)$. The interatomic distance $d$ can be estimated by the free volume following $d = \frac{1}{2}(V_0 + V_f)^{1/3}$ [17] and the diffusion limit is calculated as $u_{limit} = \frac{1}{6}|\vec{u}(\tau_0)|$ in which $\tau_0$ is the relaxation time of atomic movements, i.e., $\langle \vec{v}(\tau_0)\vec{v}(0)\rangle = 0$.

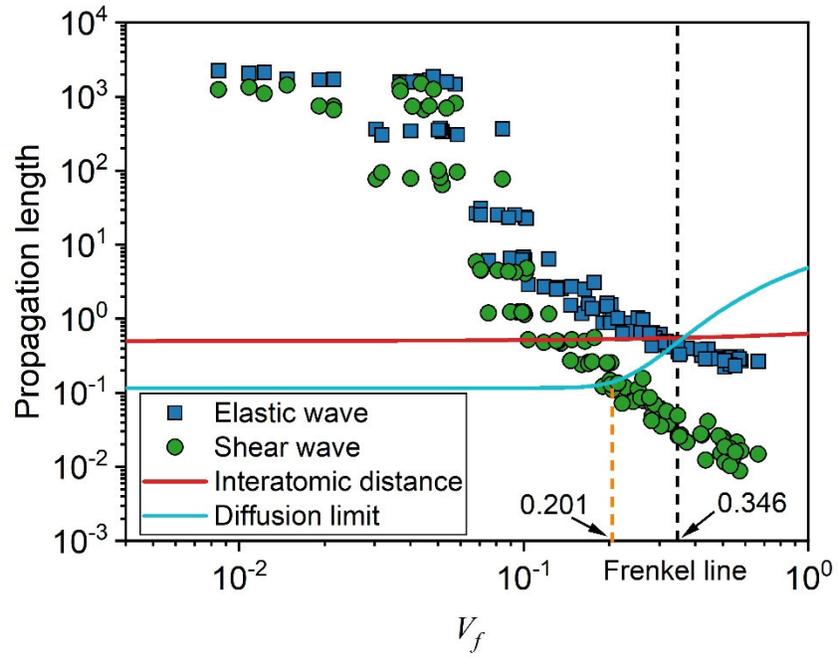

**Figure 4** The propagation length of elastic and shear waves, which are calculated as $\Lambda = v_g \tau$, in which the relaxation time of elastic and shear waves are calculated using $\tau_\vartheta = \vartheta / \vartheta^\infty$ and $\tau_\eta = \eta / \eta^\infty$, respectively.